\documentstyle[11pt]{article}
\def\rf#1{(\ref{eq:#1})}
\def\lab#1{\label{eq:#1}}
\def\nonu{\nonumber}
\def\br{\begin{eqnarray}}
\def\er{\end{eqnarray}}
\def\be{\begin{equation}}
\def\ee{\end{equation}}

\def\foot#1{\footnotemark\footnotetext{#1}}
\def\lb{\lbrack}
\def\rb{\rbrack}

\def\llb{\left\lbrack}
\def\rrb{\right\rbrack}

\def\lcurl{\left\{}
\def\rcurl{\right\}}
\def\({\left(}
\def\){\right)}
\def\lskip{\vskip\baselineskip\vskip-\parskip\noindent}
\def\mskp{\par\vskip 0.3cm \par\noindent}
\def\sskp{\par\vskip 0.15cm \par\noindent}
\def\bc{\begin{center}}
\def\ec{\end{center}}

\newcommand\partder[2]{{{\partial {#1}}\over{\partial {#2}}}}

\newcommand\sbr[2]{\left\lbrack\,{#1}\, ,\,{#2}\,\right\rbrack} 
\newcommand\Sbr[2]{\Bigl\lbrack\,{#1}\, ,\,{#2}\,\Bigr\rbrack} 
 
\def\a{\alpha}
\def\b{\beta}

\def\d{\delta}

\def\vareps{\varepsilon}

\def\h{{1\over 2}}

\def\l{\lambda}

\def\m{\mu}

\def\o{\over}

\def\vp{\varphi}
\def\P{\Phi}
\def\pa{\partial}

\def\bpa{{\bar \partial}}
\def\pr{\prime}

\def\t{\tau}

\def\wti{\widetilde}

\newcommand\sumi[1]{\sum_{#1}^{\infty}}   

\def\cA{{\cal A}}
\def\cB{{\cal B}}

\def\cF{{\cal F}}
\def\cG{{\cal G}}

\def\cL{{\cal L}}
\def\cM{{\cal M}}

\def\cX{{\cal X}}

\font \msb=msbm10 scaled \magstep1

\newcommand{\IZ}{\mbox{\msb Z} }

\def\one{\hbox{{1}\kern-.25em\hbox{l}}}

\newcommand\DB{{Darboux-B\"{a}cklund}~}
\def\Res{{\rm Res}}
\def\bt{{\bar t}}
\def\pai{\partial^{-1}\!\!}

\newcommand\st[2]{\stackrel{(#1 )}{#2}}

\newcommand\stta[2]{\stackrel{(#1 )}{t_{#2}}}

\newcommand{\ct}[1]{\cite{#1}}
\newcommand{\bi}[1]{\bibitem{#1}}
\newcommand\CMP[3]{{\sl Commun. Math. Phys.} {\bf #1} (#2) #3}

\newcommand\PLA[3]{{\sl Phys. Lett.} {\bf #1A} (#2) #3}

\newcommand\LMP[3]{{\sl Letters in Math. Phys.} {\bf #1} (#2) #3}
\newcommand\IJMPA[3]{{\sl Int. J. Mod. Phys.} {\bf A#1} (#2) #3}

\newcommand\PHSA[3]{{\sl Physica} {\bf A#1} (#2) #3}

\newcommand\JGP[3]{{\sl J. Geom. Phys.} {\bf #1} (#2) #3}

\begin{document}

\begin{center}
{\large {\bf Loop-Algebra and Virasoro Symmetries of Integrable Hierarchies of
KP Type}}
\end{center}

\begin{center}
H. Aratyn${}^1$\footnotetext[1]{Department of Physics, University of 
Illinois at Chicago, 845 W. Taylor St., Chicago, IL 60607-7059, U.S.A.;
e-mail: aratyn@uic.edu}, 
J.F. Gomes${}^{2}$\footnotetext[2]{Instituto de F\'\i sica Te\'orica -- 
IFT/UNESP, Rua Pamplona 145, 01405-900, S\~ao Paulo - SP, Brazil;
e-mail: jfg@ift.unesp.br},  
E. Nissimov${}^{3}$ and S. Pacheva${}^{3}$
\footnotetext[3]{Institute of Nuclear Research and Nuclear Energy,
Boul. Tsarigradsko Chausee 72, BG-1784 $\;$Sofia, Bulgaria;
e-mail: nissimov@inrne.bas.bg , svetlana@inrne.bas.bg}
\end{center} 

\begin{abstract}
We propose a systematic treatment of symmetries of KP integrable systems,
including constrained (reduced) KP models ${\sl cKP}_{R,M}$ (generalized AKNS 
hierarchies), and their multi-component (matrix) generalizations. Any 
${\sl cKP}_{R,M}$ integrable hierarchy is shown to possess
$\({\widehat U}(1)\oplus{\widehat {SL}}(M)\)_{+} \oplus 
\({\widehat {SL}}(M+R)\)_{-}$ loop-algebra
(additional) symmetry. Also we provide a systematic
construction of the full algebra of Virasoro additional symmetries in the case
of constrained KP models which requires a nontrivial modification of the known
Orlov-Schulman construction for the general unconstrained KP hierarchy.
Multi-component KP hierarchies are identified as ordinary (scalar)
one-component KP hierarchies supplemented with the Cartan subalgebra
of the additional symmetry algebra, which provides the basis of a
new method for construction of soliton-like solutions.
Davey-Stewartson and $N$-wave resonant systems arise as symmetry flows of
ordinary ${\sl cKP}_{R,M}$ hierarchies.
\end{abstract}

\noindent
{\bf 1. Introduction}
\sskp
The Kadomtsev-Petviashvili (KP) hierarchy of integrable soliton
evolution equations, together with its reductions and multi-component
(matrix) generalizations, describe a variety of physically important 
nonlinear phenomena (for a review, see {\sl e.g.} \ct{KP-gen,Dickey-book}).
Constrained (reduced) KP models are intimately connected with the matrix
models in non-perturbative string theory of elementary particles at
ultra-high energies (\ct{matrix-models} and references therein). They provide
an unified description of a number of basic soliton equations such as
Korteveg-de-Vries, nonlinear Schr\"{o}dinger (AKNS hierarchy in general),
Yajima-Oikawa, coupled Boussinesq-type equations etc.. Recently it was found
\ct{Wiegmann-cKP} that dispersionless KP models play a fundamental role in the
description of interface dynamics (Laplacian growth problem).
Multi-component (matrix) KP hierarchies, in turn, contain such physically
interesting systems as 2-dimensional Toda lattice, Davey-Stewartson,
$N$-wave resonant system etc.. Recently it has been shown \ct{Van-de-Leur}
that multi-component KP tau-functions provide solutions to the basic
Witten-Dijkgraaf-Verlinde-Verlinde equations in topological field theory.

In the present paper we propose a systematic approach, within Sato 
pseudo-differential operator framework, for treating symmetries of 
KP integrable systems, including constrained KP models ${\sl cKP}_{R,M}$
(generalized AKNS hierarchies -- see Eq.\rf{Lax-R-M} below), and their 
multi-component generalizations. Any ${\sl cKP}_{R,M}$ hierarchy is 
shown to possess $\({\widehat U}(1)\oplus{\widehat {SL}}(M)\)_{+} \oplus 
\({\widehat {SL}}(M+R)\)_{-}$ 
loop-algebra (additional) symmetry generated by squared eigenfunction 
potentials. The latter subscripts $(\pm)$ indicate taking the
positive/negative-grade part of the corresponding loop algebra. The symmetry
flows generating the above two mutually commuting loop-algebra factors will
be called "positive"/"negative" for brevity.

Furthermore, we provide a systematic construction of the full 
algebra of Virasoro additional symmetries in the case of constrained KP 
models which requires a nontrivial modification of the known Orlov-Schulman 
Virasoro construction \ct{Orlov-Schulman} for the general unconstrained KP 
hierarchy.

Multi-component (matrix) KP hierarchies are identified as ordinary (scalar)
one-component KP hierarchies supplemented with a special set of commuting
additional symmetries, namely, the Cartan subalgebra of the underlying
loop algebra. This identification leads to new systematic methods of
constructing soliton-like solutions of multi-component KP hierachies by
employing the well-established techniques of \DB transformations in ordinary
one-component KP hierarchies. In particular, Davey-Stewartson \ct{DS} and 
$N$-wave resonant systems arise as symmetry flows of ordinary ${\sl cKP}_{R,M}$ 
hierarchies.

\mskp
{\bf 2. Sato Formalism for Additional Symmetries of Integrable Hierarchies}
\sskp
The general one-component (scalar) KP hierarchy is given by
a pseudo-differential\foot{In what follows the operator $D$ is 
such that $\sbr{D}{f} = \pa f = \pa f /\pa x$ and the generalized Leibniz 
rule holds: $D^n f = \sumi{j=0} {n \choose j} (\pa^j f) D^{n-j}$ with 
$n \in \IZ$. In order to avoid confusion we shall employ the following 
notations: for any (pseudo-)\-differential operator $A = \sum_k a_k D^k$ and 
a function $f$, the symbol $\, A(f)\,$ will indicate application (action) of
$A$ on $f$, whereas the symbol $Af$ will denote simply operator product of
$A$ with the zero-order (multiplication) operator $f$. Projections $(\pm)$
are defined as: $A_{+} = \sum_{k\geq 0} a_k D^k$ and 
$A_{-} = \sum_{k\leq -1} a_k D^k$. Finally, $\Res A \equiv a_{-1}$.}  
Lax operator $\cL$ obeying Sato evolution equations
(also known as isospectral flow equations; for a systematic exposition, see
\ct{Dickey-book}) :   
\be
\cL = D + \sum_{k=1}^\infty u_k D^{-k}    \quad , \quad
\partder{}{t_n}\cL = \Sbr{\(\cL^n\)_{+}}{\cL} 
\lab{Lax-gen}
\ee
with Sato dressing operator $W$ :
\be
\cL = W D W^{-1} \quad , \quad  \partder{}{t_n} W = - \(W D^n W^{-1}\)_{-} W
\quad ,\quad 
W = \sum_{k=0}^\infty \frac{p_k (-[\pa])\t (t)}{\t (t)} D^{-k}
\lab{Sato-dress}
\ee
and (adjoint) Baker-Akhiezer (BA) wave functions $\psi_{BA}^{(\ast)} (t,\l)$ :
\be
\cL^{(\ast)} \psi_{BA}^{(\ast)} = \l \psi_{BA}^{(\ast)} \quad , \quad
\partder{}{t_n} \psi_{BA}^{(\ast)} = 
\pm \({\cL^{(\ast)}}^n\)_{+} (\psi_{BA}^{(\ast)})
\lab{BA-eqs}
\ee
\be
\psi_{BA}^{(\ast)} (t,\l) = W^{(\ast\, -1)} \( e^{\pm \xi (t,\l)}\) =
\frac{\t (t \mp [\l^{-1}])}{\t (t)}\, e^{\pm \xi (t,\l)} \quad ,\quad
 \xi (t,\l) \equiv \sum_{\ell =1}^\infty t_\ell \l^{\ell}
\lab{BA-def}
\ee
where the tau-function $\t (t)$ satisfies the relation:
\be
\pa_x \partder{}{t_n} \ln \t = \Res \cL^n
\lab{tau-basic}
\ee
Here and below we employ the following short-hand notations:
$(t) \equiv (t_1 \equiv x, t_2, \ldots )$ for the set of isospectral
time-evolution parameters;
$[\pa] \equiv \Bigl(\partder{}{t_1}, \h \partder{}{t_2},
{1\o 3}\partder{}{t_3}, \ldots \Bigr)$ 
and 
$[\l^{-1}] \equiv \Bigl(\l^{-1}, \h \l^{-2}, {1\o 3} \l^{-3},\ldots \Bigr)$;
$p_k (.)$ indicate the well-known Schur polynomials.

There exist few other objects in Sato formalism for integrable hierarchies
which play fundamental role in our construction.
(Adjoint) eigenfunctions $\P (t)$ ($\Psi (t)$, respectively) are those
functions of KP ``times'' $(t)$ satisfying:
\be
\partder{}{t_l} \P = (\cL^l)_{+} (\P) \quad ,\quad
\partder{}{t_l} \Psi = - (\cL^l)^\ast_{+} (\Psi)
\lab{EF-def}
\ee
According to second Eq.\rf{BA-eqs}, (adjoint) BA functions are special
cases of (adjoint) eigenfunctions, which in addition satisfy spectral
equations (first Eq.\rf{BA-eqs}). 

It has been shown in ref.\ct{ridge} that any (adjoint) eigenfunction possesses
a spectral representation of the form\foot{Integrals over spectral parameters
are understood as: 
$\int d\l \equiv \oint_{0} \frac{d\l}{2i\pi} ={\rm Res}_{\l = 0}$.} :
\be
\P (t) = \int d\l\, \vp (\l)\, \psi_{BA}(t,\l) \quad ,\quad
\Psi (t) = \int d\l\, \psi (\l)\, \psi^\ast_{BA}(t,\l)
\lab{spec-repr}
\ee
with appropriate spectral densities $\vp (\l)$ and $\psi (\l)$ which are
formal Laurent series in $\l$. Clearly, any KP hierarchy possesses an
infinite set of independent (adjoint) eigenfunctions in one-to-one
correspondence with the space of all independent formal Laurent series in
$\l$.

The next important object is the so called squared eigenfunction potential 
(SEP) \ct{oevela} -- a function $S\(\P (t),\Psi (t)\)$ associated with 
an arbitrary pair of (adjoint) eigenfunctions $\P (t),\Psi (t)$ 
which possesses the following characteristics:
\be
{\pa \o \pa t_n} S \( \P (t) , \Psi (t)\) = 
{\rm Res} \( D^{-1} \Psi (\cL^n)_{+} \P D^{-1} \)
\lab{potentialflo}
\ee
In particular, for $n=1$ Eq.\rf{potentialflo} implies
~$\pa_x S\(\P (t),\Psi (t)\) = \P (t)\,\Psi (t)$
(recall $\pa_x \equiv \pa/\pa t_1$). Eq.\rf{potentialflo} determines
$S\(\P (t),\Psi (t)\) \equiv \pai\(\P (t)\,\Psi (t)\)$ up to a shift by a 
trivial constant which is uniquely fixed by the fact that any SEP obeys the
following double-spectral representation \ct{ridge} :
\br
\pai\(\P (t)\,\Psi (t)\) = -\int\!\int d\l d\m \, \psi (\l)\,\vp (\m) \,
{1\o \l}\psi^\ast_{BA}(t,\l) \psi_{BA} (t + [\l^{-1}],\m) 
\nonu \\
=  -\int\!\int d\l d\m \, \frac{\psi (\l)\,\vp (\m)}{\l - \m}
e^{\xi (t,\m) - \xi (t,\l)} \frac{\t (t+[\l^{-1}]-[\m^{-1}])}{\t (t)}
\lab{SEP-spec}
\er
with $\vp (\l), \,\psi (\l)$ being the respective spectral densities in
\rf{spec-repr}.
It is in this well-defined sense that inverse space derivatives
$\pai$ will appear throughout our construction below.

A flow on the space of Sato pseudo-differential Lax operators $\cL$ or,
equivalently, on the space of Sato dressing operators $W$ is given by:
\be
\d_\a \cL = \Sbr{\cM_\a}{\cL}  \quad ,\quad  \d_\a W = \cM_\a W
\lab{flow-def}
\ee
where $\cM_\a$ is a purely pseudo-differential operator. A flow 
$\d_\a$ \rf{flow-def} is a symmetry if and only if it commutes with the 
isospectral flows $\partder{}{t_l}$ :
\be
\Sbr{\d_\a}{\partder{}{t_l}} = 0  \quad \longrightarrow \quad
\partder{}{t_l}\cM_\a = {\Sbr{(\cL^l)_{+}}{\cM_\a}}_{-}
\lab{symm-def}
\ee
The general form of $\cM_\a$ obeying \rf{symm-def} is provided by
ref.\ct{Dickey-95} :
\be
\cM_\a = \int\!\int\! d\l d\m \,\rho_\a (\l,\m)\, 
\psi_{BA}(t,\m) D^{-1} \psi^\ast_{BA}(t,\l) =
\sum_{I,J \in \cA} c^{(\a)}_{IJ}\, \P_J D^{-1} \Psi_I
\lab{M-bispec}
\ee
where $\rho_\a (\l,\m)$ is arbitrary (in the case of the general
unconstrained KP hierarchy) double Laurent series in $\l$ and $\m$.
In the second equality above the sums run in general over an infinite set 
$\cA$ of indices, and $\lcurl \P_I ,\,\Psi_I \rcurl_{I \in \cA}$ are
(adjoint) eigenfunctions of the Lax operator $\cL$ \rf{EF-def}
for $\P = \P_I$ and $\Psi = \Psi_I$.
The second equality in \rf{M-bispec} arises from the general representation
of the ``bispectral'' density:
\be
\rho_\a (\l,\m) = \sum_{I,J \in \cA} c^{(\a)}_{IJ}\, 
\vp_J (\m) \psi_I (\l)
\lab{bispec-repr}
\ee
in terms of basis of Laurent series $\lcurl \vp_I (\m)\rcurl$ and
$\lcurl \psi_I (\l)\rcurl$, together with spectral representation
theorem \ct{ridge} for (adjoint) eigenfunctions of Sato Lax operators
(cf. Eqs.\rf{spec-repr}). 

Our main objective below will be to construct explicitly symmetry-flow
generating operators \rf{M-bispec}, such that the corresponding flows
both preserve the constrained
form of Lax operators definining constrained KP hierarchies, as well as they
yield a closed algebra of symmetries. At this point let us recall that the 
following special form of 
$\rho_\a (\l,\m) = \l^k \bigl(\pa/\pa \l\bigr)^l \d (\l - \m)$
in \rf{M-bispec} \ct{Dickey-95} yields the well-known Orlov-Schulman 
$W_{1+\infty}$ additional symmetries \ct{Orlov-Schulman} in the case of the 
general unconstrained KP hierarchy. On the other hand, these standard 
Orlov-Schulman symmetry flows fail to produce symmetries in the case of 
constrained KP hierarchies since they do not preserve the constrained form of 
the pertinent Lax operators. The solution to this problem is provided by a 
non-trivial modification of the former flows (see ref.\ct{noak-addsym} 
and Section 6 below).

On any (adjoint) eigenfunction the action of the flow $\d_\a$ \rf{flow-def}
takes the form:
\be
\d_\a \P = \cM_\a (\P) + \cF^{(\a)}  \quad ,\quad
\d_\a \Psi = - \cM^\ast_\a (\Psi) + \cG^{(\a)}
\lab{flow-EF-def}
\ee
where $\cF^{(\a)}$ and $\cG^{(\a)}$ are other (adjoint) eigenfunctions.
Eqs.\rf{flow-EF-def} follow from the second Eq.\rf{flow-def} taking again
into account the spectral representation theorem \ct{ridge}. Note that the
emergence of the additional (adjoint) eigenfunctions terms on the r.h.s. of
\rf{flow-EF-def} is due to the fact that the spectral densities of $\P$ and
$\Psi$ in \rf{spec-repr} may in general vary under the action of $\d_\a$.
Moreover, as it will be seen in Section 4 below, in the case of
constrained KP hierarchies the presence of the additional terms on the r.h.s.
of Eqs.\rf{flow-EF-def} is mandatory for consistency of the flow action 
\rf{flow-def} with the constrained form of the pertinent Lax operator,
which accordingly uniquely fixes the form of $\cF^{(\a)}$ and $\cG^{(\a)}$.

Making use of the well-known pseudo-differential operator identities (cf. 
{\sl e.g.} the appendix in first ref.\ct{noak-addsym}) :
\br
M_1 M_2 = M_1 (f_2) D^{-1} g_2 + f_1 D^{-1} M_2^\ast (g_1) 
\lab{pseudo-diff-id}  \\
M_{1,2} \equiv f_{1,2} D^{-1} g_{1,2} \quad ,\quad
M_1 (f_2) = f_1 \pai\(g_1 f_2\) \;\; etc.
\nonu
\er
one easily finds that the flows $\d_a$ \rf{flow-def}
span (an infinite-dimensional, in general) closed algebra:
\be
\Sbr{\d_\a}{\d_\b} \cL = \Sbr{\d_\a \cM_\b - \d_\b \cM_\a -
\lb \cM_\a,\, \cM_\b\rb}{\cL}
\lab{commut-flows}
\ee
where:
\br
\d_\a \cM_\b - \d_\b \cM_\a - \lb \cM_\a,\, \cM_\b\rb  
\equiv \cM_{\lb\a ,\b\rb} =
\nonu \\
\sum_{I,J \in \cA} \( c^{(\b)} a^{(\a)} - c^{(\a)} a^{(\b)}
+ b^{(\a)} c^{(\b)} - b^{(\b)} c^{(\a)}\)_{IJ} \P_J D^{-1}\Psi_I
\lab{M-commut-def}
\er
Here the matrices $a^{(\a)}_{IJ}$ and $b^{(\a)}_{IJ}$ appear in the
inhomogeneous terms in the $\d_\a$-flow equations for $\P_I$ and
$\Psi_I$, respectively, according to Eqs.\rf{flow-EF-def} :
\be
\d_\a \P_I = \cM_\a (\P_I) + \sum_J a^{(\a)}_{IJ} \P_J  \quad ,\quad
\d_\a \Psi_I = - \cM^\ast_\a (\Psi_I) + \sum_J b^{(\a)}_{JI} \Psi_J
\lab{flow-EF-I}
\ee
and similarly for $a^{(\b)}_{IJ}$ and $b^{(\b)}_{IJ}$.
In the case of general KP hierarchy
the specific form of $a^{(\a)}_{IJ}$ and $b^{(\a)}_{IJ}$ for any flow 
(with label $\a$) is arbitrary {\sl a priori}, and it is subject to the only
condition of fulfillment of Jacobi identities for the flow commutator
\rf{M-commut-def}. However, for constrained KP hierarchies the form of
$a^{(\a)}_{IJ}$ and $b^{(\a)}_{IJ}$ is determined uniquely from the
consistency (Eq.\rf{consist-cond} below) of the flow action with the 
constrained form of the pertinent Lax operator, see Section 4 below.

Finally, starting from relation \rf{tau-basic} and using \rf{symm-def}
we find for the transformation of the tau-function under the action of
$\d_\a$-flow \rf{flow-def} :
\be
\d_\a \ln \t = - \pai\(\Res \cM_\a\) =
- \sum_{I,J} c^{(\a)}_{IJ} \pai\(\P_J \Psi_I\)
\lab{tau-flow-a}
\ee
\mskp
{\bf 3. Constrained KP Hierarchies. Inverse Powers of Lax Operators}
\sskp
So far we have considered the general case of unconstrained KP hierarchy.
Now we are interested in symmetries for {\em constrained} KP hierarchies
${\sf cKP}_{R,M}$ with Lax operators (cf. 
\ct{noak-addsym,ridge,UIC-97} and references therein) :
\be
\cL \equiv \cL_{R,M} = D^{R} + \sum_{i=0}^{R-2} v_i D^i + 
\sum_{j=1}^M \P_j D^{-1} \Psi_j = L_{M+R} L_M^{-1}
\lab{Lax-R-M}
\ee
where $\lcurl \P_i,\,\Psi_i\rcurl_{i=1}^M$ is a set of (adjoint)
eigenfunctions of $\cL$.

The second representation of $\cL \equiv \cL_{R,M}$\foot{Henceforth we shall
employ the short-hand notation $\cL$ for $\cL_{R,M}$ \rf{Lax-R-M} whenever
this will not lead to a confusion.}
is in terms of a ratio of two monic 
purely differential operators $L_{M+R}$ and $L_M$ of orders $M+R$ 
and $M$, respectively (see \ct{UIC-97} and references therein). For 
$\cL \equiv \cL_{R,M}$ the Sato evolution (isospectral flow) Eqs.\rf{Lax-gen},
the equations for (adjoint) BA \rf{BA-eqs} and (adjoint) eigenfunctions
\rf{EF-def} acquire the form:
\be
\partder{}{t_n}\cL = \Sbr{\(\cL^{n\o R}\)_{+}}{\cL} \quad ,\quad
\cL^{(\ast)} \psi_{BA}^{(\ast)} = \l^R \psi_{BA}^{(\ast)} \quad , \quad
\partder{}{t_n} \psi_{BA}^{(\ast)} = 
\pm (\cL^{(\ast)})^{n\o R}_{+} (\psi_{BA}^{(\ast)})
\lab{Lax-BA-eqs-R}
\ee
\be
\partder{}{t_n} \P = (\cL^{n\o R})_{+} (\P) \quad ,\quad
\partder{}{t_n} \Psi = - (\cL^{n\o R})^\ast_{+} (\Psi)
\lab{EF-def-R}
\ee

In the case of constrained hierarchies  \rf{Lax-R-M}, we have the
following additional condition on the symmetry generating operator $\cM_\a$ 
since the flow \rf{flow-def} must preserve the constrained form 
\rf{Lax-R-M} of the pertinent Lax operator (cf.\rf{flow-EF-def} and 
\rf{M-bispec}) :
\be
\sum_{i=1}^M \llb \cF^{(\a)}_i D^{-1} \Psi_i + \P_i D^{-1} \cG^{(\a)}_i \rrb
= \sum_{I,J \in \{\a\}} c_{IJ}\, 
\llb \P^{(\a)}_J D^{-1} \cL^\ast (\Psi^{(\a)}_I) -
\cL (\P^{(\a)}_J) D^{-1} \Psi^{(\a)}_I \rrb 
\lab{consist-cond}
\ee
where (cf. \rf{flow-EF-def}) :
\be
\d_\a \P_i = \cM_\a (\P_i) + \cF^{(\a)}_i  \quad ,\quad
\d_\a \Psi_i = - \cM^\ast_\a (\Psi_i) + \cG^{(\a)}_i
\lab{flow-EF-i-def}
\ee
Eq.\rf{consist-cond} uniquely fixes the form of the additional terms 
$\cF^{(\a)}_i$ and $\cG^{(\a)}_i$ in \rf{flow-EF-i-def}.

In what follows we will also need the $\d_\a$-flow equations on inverse
powers of the Lax operator $\cL = L_{M+R} L^{-1}_M$ \rf{Lax-R-M}.
First, let us recall that the inverses of the underlying purely differential
operators are given by:
\be
L_M^{-1} = \sum_{i=1}^M \vp_i D^{-1} \psi_i  \quad ,\quad
L_{M+R}^{-1} = \sum_{a=1}^{M+R} {\bar {\vp}}_a D^{-1} {\bar {\psi}}_a
\lab{inverses}
\ee
where the functions $\lcurl \vp_i \rcurl_{i=1}^M$ and 
$\lcurl \psi_i \rcurl_{i=1}^M$ span $Ker (L_M)$ and $Ker(L_M^\ast)$, 
respectively, whereas $\lcurl {\bar {\vp}}_a \rcurl_{a=1}^{M+R}$ and
$\lcurl {\bar {\psi}}_a \rcurl_{a=1}^{M+R}$ span $Ker (L_{M+R})$ and 
$Ker(L_{M+R}^\ast)$, respectively. Therefore we have:
\be
\cL = (\cL )_{+} + \sum_{i=1}^M L_{M+R}(\vp_i) D^{-1} \psi_i 
\quad ,\quad i.e. \;\;
\P_i =  L_{M+R}(\vp_i) \;\; ,\;\; \Psi_i = \psi_i
\lab{Lax-plus-1}
\ee
\be
\cL^{-1} = \sum_{a=1}^{M+R} L_M ({\bar {\vp}}_a) D^{-1} {\bar {\psi}}_a
\lab{L-minus-1}
\ee
\be
\cL^{-N} = \sum_{a=1}^{M+R} \sum_{s=0}^{N-1}
\cL^{-(N-1)+s} \bigl( L_M ({\bar {\vp}}_a)\bigr) D^{-1} 
\Bigl(\cL^{-s}\Bigr)^\ast ({\bar {\psi}}_a)
\lab{L-minus-N}
\ee
Compare the last formula \rf{L-minus-N} with the formula \ct{EOR-95} for the 
negative pseudo-differential part of a positive power of $\cL$ \rf{Lax-R-M}:
\be
\Bigl(\cL^N\Bigr)_{-} = \sum_{i=1}^M \sum_{s=0}^{N-1}
\cL^{N-1-s}(\P_i) D^{-1} \Bigl(\cL^s\Bigr)^\ast (\Psi_i)
\lab{L-plus-N}
\ee

Let us also note that the following simple consequences from the 
definitions of the corresponding objects will play essential role for the 
consistency of the constructions involving inverse powers of $\cL$ :
\be
\cL \bigl( L_M ({\bar{\vp}}_a)\bigr) = 0  \quad ,\quad
\cL^\ast ({\bar {\psi}}_a) = 0   \quad ,\quad
\cL^{-1} (\P_i) = 0 \quad ,\quad  \(\cL^{-1}\)^\ast (\Psi_i) = 0
\lab{zero-eqs}
\ee

Applying the flow Eq.\rf{flow-def} to $\cL^{-1}$ \rf{L-minus-1}
$\d_\a \cL^{-1} = \Sbr{\cM_\a}{\cL^{-1}}$ and taking into account the explicit 
form of $\cM_\a$ (second equality \rf{M-bispec}) we obtain:
\be
\d_\a \bigl( L_M ({\bar{\vp}}_a)\bigr) = \cM_\a \bigl( L_M ({\bar{\vp}}_a)\bigr)
+ {\bar \cF}^{(\a)}_a  \quad ,\quad
\d_\a {\bar {\psi}}_a = - \cM^\ast_\a ({\bar {\psi}}_a)+ {\bar \cG}^{(\a)}_a
\lab{flow-EF-inverse}
\ee
with consistency condition for the ``shift'' functions ${\bar \cF}^{(\a)}_a$
and ${\bar \cG}^{(\a)}_a$ (the analog of Eq.\rf{flow-EF-i-def}) :
\be
\sum_{a=1}^{M+R} \llb {\bar \cF}^{(\a)}_a D^{-1} {\bar {\psi}}_a + 
L_M ({\bar{\vp}}_a) D^{-1} {\bar \cG}^{(\a)}_a \rrb
= \sum_{I,J \in \cA} c^{(\a)}_{IJ}\llb \P_J D^{-1} \(\cL^{-1}\)^\ast (\Psi_I)
- \cL^{-1} (\P_J) D^{-1} \Psi_I \rrb 
\lab{consist-cond-inverse}
\ee

Also, from the isospectral flows equations applied on $\cL^{-1}$, 
{\sl i.e.}, $\partder{}{t_n} \cL^{-1} = \Sbr{\cL^{n\o R}_{+}}{\cL^{-1}}$, we
find, taking into account \rf{zero-eqs}, that $L_M ({\bar{\vp}}_a)$ and 
${\bar {\psi}}_a$ are (adjoint) eigenfunctions of $\cL$ (cf. \rf{EF-def-R}) :
\be
\partder{}{t_n} L_M ({\bar{\vp}}_a)  = (\cL^{n\o R})_{+} 
\bigl( L_M ({\bar{\vp}}_a)\bigr) \quad ,\quad
\partder{}{t_n} {\bar {\psi}}_a = - (\cL^{n\o R})^\ast_{+} ({\bar {\psi}}_a)
\lab{EF-def-inverse}
\ee
\mskp
{\bf 4. Loop-Algebra Symmetries of KP Hierarchies}
\sskp
Let us consider the following system of $M$ infinite sets of (adjoint)
eigenfunctions of $\cL \equiv \cL_{R,M}$ \rf{Lax-R-M} :
\be
\P^{(n)}_i \equiv \cL^{n-1}(\P_i) \quad ,\quad
\Psi^{(n)}_i \equiv \(\cL^\ast\)^{n-1}(\Psi_i) \quad ,\;\;\; n=1,2,\ldots \; ;
\;\; i=1,\ldots ,M
\lab{EF-cKP-sys}
\ee
which are expressed in terms of the $M$ pairs of (adjoint) eigenfunctions
entering the pseudo-differential part of $\cL \equiv \cL_{R,M}$ \rf{Lax-R-M}.
Using \rf{EF-cKP-sys} we can build the following infinite set of symmetry 
flows (cf. \rf{flow-def} and \rf{M-bispec}) :
\be
\d^{(n)}_A \cL = \Sbr{\cM^{(n)}_A}{\cL}       \quad ,\quad
\cM^{(n)}_A \equiv \sum_{i,j=1}^M A^{(n)}_{ij} 
\sum_{s=1}^n \P^{(n+1-s)}_j D^{-1} \Psi^{(s)}_i
\lab{flow-n-A}
\ee
where $A^{(n)}$ is an arbitrary constant $M \times M$ matrix, {\sl i.e.},
$A^{(n)} \in Mat(M)$. 
Consistency of the flow action \rf{flow-n-A} with the constrained form
\rf{Lax-R-M} of $\cL \equiv \cL_{R,M}$ (cf. \rf{consist-cond}) implies the 
following flow action on the involved (adjoint)
eigenfunctions: 
\br
\d^{(n)}_A \P^{(m)}_i = \cM^{(n)}_A (\P^{(m)}_i) - 
\sum_{j=1}^M A^{(n)}_{ij}\P^{(n+m)}_j 
\nonu \\
\d^{(n)}_A \Psi^{(m)}_i = -\(\cM^{(n)}_A\)^\ast (\Psi^{(m)}_i) + 
\sum_{j=1}^M A^{(n)}_{ji}\Psi^{(n+m)}_j 
\lab{flow-n-A-EF}
\er
The specific form of the inhomogeneous terms on the r.h.s. of 
Eqs.\rf{flow-n-A-EF} is the main ingredient of our construction. It is
precisely these inhomogeneous terms which yield non-trivial loop-algebra
additional symmetries.

Using the pseudo-differential operator identities \rf{pseudo-diff-id}
and taking into account \rf{flow-n-A-EF} we can show that 
(cf. Eq.\rf{M-commut-def}):
\be
\d^{(n)}_A \cM^{(m)}_B - \d^{(m)}_B \cM^{(n)}_A - 
\Sbr{\cM^{(n)}_A}{\cM^{(m)}_B} = \cM^{(n+m)}_{\lb A,\, B\rb}
\lab{M-commut-A-B}
\ee
Eq.\rf{M-commut-A-B} implies that the symmetry flows \rf{flow-n-A}--\rf{flow-n-A-EF}
span the following infinite-dimensional algebra:
\be
\Sbr{\d^{(n)}_A}{\d^{(m)}_B} = \d^{(n+m)}_{\lb A,\, B\rb} \quad ;\quad
A^{(n)},B^{(m)} \in Mat(M) \;\; ,\;\; n,m =1,2, \ldots
\lab{KM-alg-flows}
\ee
isomorphic to $\({\widehat U}(1) \times {\widehat {SL}}(M)\)_{+}$ where the
subscript $(+)$ indicates taking the positive-grade subalgebra of the
corresponding loop-algebra.
We observe, that in the case of ${\sf cKP}_{R,M}$ models we have
~$\cM^{(n)}_{A=\one} = \(\cL_{R,M}^n\)_{-}$ (insert \rf{EF-cKP-sys} into
first relation \rf{flow-n-A} for $A^{(n)}=\one$ and compare with \rf{L-plus-N}).
Therefore, the flows $\d^{(n)}_{A=\one}$ for ${\sf cKP}_{R,M}$ models
coincide upto a sign with the ordinary isospectral flows modulo $R$:
$\d^{(n)}_{A=\one} = -\partder{}{t_{nR}}$ (cf. Eqs.\rf{Lax-BA-eqs-R}). 
Thereby the flows $\d^{(n)}_A$ \rf{flow-n-A} will be called "positive" for brevity.


Now we consider another infinite set of (adjoint) eigenfunctions of
$\cL \equiv \cL_{R,M}$ expressed in terms of the (adjoint) eigenfunctions
entering the inverse power of $\cL \equiv \cL_{R,M}$ \rf{L-minus-1} :
\br
\P^{(-m)}_a \equiv \cL^{-(m-1)}\bigl( L_M ({\bar \vp}_a)\bigr) \quad ,\quad
\Psi^{(-m)}_a \equiv \(\cL^{-(m-1)}\)^\ast ({\bar \psi}_a) 
\lab{EF-sys-inverse} \\
m=1,2,\ldots \; , \;\; a=1,\ldots , M+R
\er
Using \rf{EF-sys-inverse} we obtain the following set of "negative" symmetry 
flows which parallels completely the set of "positive" flows \rf{flow-n-A} :
\be
\d^{(-n)}_\cA \cL = \Sbr{\cM^{(-n)}_\cA}{\cL}     \quad ,\quad
\cM^{(-n)}_\cA \equiv \sum_{a,b=1}^{M+R} \cA^{(-n)}_{ab} 
\sum_{s=1}^n \P^{(-n-1+s)}_b D^{-1} \Psi^{(-s)}_a
\lab{flow-n-A-neg}
\ee
where $\cA^{(-n)}_{ab}$ is an arbitrary constant $(M+R) \times (M+R)$ matrix, 
{\sl i.e.}, $\cA^{(-n)} \in Mat(M+R)$. In fact, since according to 
\rf{L-minus-N} we have $\cM^{(-n)}_{\cA = \one}= \cL^{-n}$, the flows 
$\d^{(-n)}_{\cA = \one}$ vanish identically, {\sl i.e.},
$\d^{(-n)}_{\cA = \one} = 0$, therefore, we restrict $\cA^{(-n)} \in SL(M+R)$.

Consistency of the flow action \rf{flow-n-A-neg} with the constrained form
\rf{Lax-R-M} of $\cL \equiv \cL_{R,M}$ (cf. \rf{consist-cond}) 
and with the constrained form \rf{L-minus-1} of the inverse $\cL^{-1}$
implies the following $\d^{(-n)}_\cA$-flow action on the involved (adjoint)
eigenfunctions (using short-hand notations \rf{EF-cKP-sys} and 
\rf{EF-sys-inverse}) :
\be
\d^{(-n)}_\cA \P^{(m)}_i = \cM^{(-n)}_\cA (\P^{(m)}_i)    \quad ,\quad
\d^{(-n)}_\cA \Psi^{(m)}_i = -\(\cM^{(-n)}_\cA\)^\ast (\Psi^{(m)}_i)
\lab{flow-n-A-neg-EF}
\ee
\br
\d^{(-n)}_\cA \P^{(-m)}_a = \cM^{(-n)}_\cA (\P^{(-m)}_a) - 
\sum_{b=1}^{M+R} \cA^{(-n)}_{ab}\P^{(-n-m)}_b 
\nonu \\
\d^{(-n)}_\cA \Psi^{(-m)}_a = -\(\cM^{(-n)}_\cA\)^\ast (\Psi^{(-m)}_a) + 
\sum_{b=1}^{M+R} \cA^{(-n)}_{ba}\Psi^{(-n-m)}_b 
\lab{flow-n-A-neg-EF-inverse}
\er
Similarly, consistency of "positive" $\d^{(n)}_A$-flow action \rf{flow-n-A}
with the constrained form \rf{L-minus-1} of the inverse Lax operator implies:
\be
\d^{(n)}_A \P^{(-m)}_a = \cM^{(n)}_A (\P^{(-m)}_a)   \quad ,\quad
\d^{(n)}_A \Psi^{(-m)}_a = -\(\cM^{(n)}_A\)^\ast (\Psi^{(-m)}_a)
\lab{flow-n-A-EF-inverse}
\ee
Using again the pseudo-differential operator identities \rf{pseudo-diff-id}
we find from \rf{flow-n-A-neg-EF}--\rf{flow-n-A-EF-inverse} 
(cf. Eq.\rf{M-commut-A-B}) :
\be
\d^{(n)}_A \cM^{(-m)}_\cB - \d^{(-m)}_\cB \cM^{(n)}_A -
\Sbr{\cM^{(n)}_A}{\cM^{(-m)}_\cB} = 0
\lab{M-commut-A-B-neg-plus}
\ee
\be
\d^{(-n)}_\cA \cM^{(-m)}_\cB - \d^{(-m)}_\cB \cM^{(-n)}_\cA - 
\Sbr{\cM^{(-n)}_\cA}{\cM^{(-m)}_\cB} = \cM^{(-n-m)}_{\lb \cA,\,\cB\rb}
\lab{M-commut-A-B-neg}
\ee
Eqs.\rf{M-commut-A-B-neg-plus}--\rf{M-commut-A-B-neg} imply that the 
"negative" symmetry flows \rf{flow-n-A-neg}--\rf{flow-n-A-neg-EF-inverse} 
commute with the "positive" flows \rf{flow-n-A}--\rf{flow-n-A-EF}:
\be
\Sbr{\d^{(n)}_A}{\d^{(-m)}_\cB} = 0
\lab{plus-neg-commut}
\ee
and that they themselves span the following infinite-dimensional algebra:
\be
\Sbr{\d^{(-n)}_\cA}{\d^{(-m)}_\cB} = \d^{(-n-m)}_{\lb \cA,\,\cB\rb} 
\quad ;\quad
\cA^{(-n)},\cB^{(-m)} \in SL(M+R) \;\; ,\;\; n,m =1,2, \ldots
\lab{KM-alg-neg-flows}
\ee
which is isomorphic to $\({\widehat {SL}}(M+R)\)_{-}$ (the
subscript $(-)$ indicates taking the negative-grade subalgebra of the
corresponding loop-algebra). 

Therefore,we conclude that the full loop algebra of (additional) symmetries of
${\sl cKP}_{R,M}$ hierarchies \rf{Lax-R-M} is the direct sum:
\be
\({\widehat U}(1)\oplus{\widehat {SL}}(M)\)_{+} \oplus \({\widehat {SL}}(M+R)\)_{-}
\lab{KM-alg-full}
\ee

The construction above can be straightforwardly extended to the case of the
general unconstrained KP hierarchy defined by \rf{Lax-gen}. All relations
\rf{flow-n-A}--\rf{KM-alg-flows} and
\rf{flow-n-A-neg}--\rf{KM-alg-neg-flows} remain intact where now:
\be
\lcurl \P^{(n)}_i ,\,\Psi^{(n)}_i \rcurl^{n=1,2,\ldots}_{i=1,\ldots ,M}
\quad ,\quad
\lcurl \P^{(-n)}_a ,\,\Psi^{(-n)}_a \rcurl^{n=1,2,\ldots}_{a=1,\ldots ,M+R}
\lab{EF-sys}
\ee
form an infinite system of independent (adjoint) eigenfunctions of the general
Lax operator \rf{Lax-gen} with $M$, $M+R$ being arbitrary positive integers.
\lskip
{\bf 5. Multi-Component KP Hierarchies from One-Component Ones}
\sskp
Let us now consider the following subset of "positive" flows $\d^{(n)}_{E_k}$ 
\rf{flow-n-A} for the general KP hierarchy \rf{Lax-gen} corresponding to:
\be
E_k = diag(0,\ldots 0,1,0, \ldots, 0) \quad , \quad i.e. \quad
\cM^{(n)}_{E_k} = \sum_{s=1}^n \P^{(n+1-s)}_k D^{-1} \Psi^{(s)}_k
\lab{ghost-k}
\ee
Due to Eq.\rf{KM-alg-flows} the flows $\d^{(n)}_{E_k}$ span an 
infinite-dimensional Abelian algebra and, by construction, they commute with 
the original isospectral flows $\partder{}{t_n}$ as well. 
Comparison with our construction in
refs.\ct{multi-comp-KP} allows us to identify the set of isospectral flows
plus the set of $\d^{(n)}_{E_k}$-flows \rf{ghost-k} :
\be
\partder{}{t_n} \equiv \pa/\pa \stta{1}{n} \quad , \quad
\d^{(n)}_{E_k} \equiv \pa/\pa\!\!\stta{k+1}{n} \quad ,\quad k=1,\ldots ,M
\lab{multi-comp-KP-flows}
\ee
with the set of isospectral flows
$\Bigl\{\stta{\ell}{n}\Bigr\}^{\ell =1,\ldots ,M+1}_{n=1,2,\ldots}$ of the
(unconstrained) $M+1$-component matrix KP hierarchy. The latter is defined
in terms of the $M+1 \times M+1$ 
matrix Hirota bilinear identities (see refs.\ct{multi-comp-KP}) :
\br
\sum_{k=1}^{M+1} \vareps_{ik}\, \vareps_{jk}\int\! d\l \,
\l^{\d_{ik} + \d_{jk} -2}\, e^{\xi (\st{k}{t}-\st{k}{t^\pr},\l)}
\t_{ik}\bigl(\ldots ,\st{k}{t}-[\l^{-1}],\ldots \bigr)\,
\t_{kj}\bigl(\ldots ,\st{k}{t^\pr}+[\l^{-1}],\ldots \bigr) = 0
\lab{HBI}
\er
which are obeyed by a set of $M(M+1)+1$ tau-functions $\lcurl \t_{ij}\rcurl$
expressed in terms of the single tau-function $\t$ and the ``positive''
symmetry flow generating (adjoint) eigenfunctions \rf{EF-sys} in the original 
one-component (scalar) KP hierarchy \rf{Lax-gen}--\rf{tau-basic} as follows:
\be
\t_{11}=\t_{ii} = \t \quad ,\quad
\t_{1i} = \t\,\P^{(1)}_{i-1} \quad ,\quad \t_{i1} = - \t\,\Psi^{(1)}_{i-1}
\nonu
\ee
\be
\t_{ij} = \vareps_{ij} \t\,\pai\(\P^{(1)}_{j-1}\Psi^{(1)}_{i-1}\)
\quad ,\quad  i \neq j \;\; ,\;\; i,j = 2, \ldots M+1
\lab{tau-M+1-comp}
\ee
Here $\vareps_{ij} =1$ for $i \leq j$ and $\vareps_{ij} =-1$ for $i>j$, and
$\d_{ij}$ are the usual Kronecker symbols.

The above construction of multi-component (matrix) KP hierarchies out of
ordinary one-component ones can be straighforwardly carried over to the case of
constrained KP models \rf{Lax-R-M} using the identification
\rf{EF-cKP-sys} for the symmetry-generating (adjoint) eigenfunctions.
In this case, however, there is a linear dependence among the flows
\rf{multi-comp-KP-flows}
~$\sum_{k=1}^M \d^{(n)}_{E_k} = - \partder{}{t_n}$, therefore, the associated
constrained multi-component KP hierarhy is now $M\times M$ matrix hierarchy.

Similarly, we can start with the subset of ``negative'' symmetry flows
$\d^{(-n)}_{E_k}$ \rf{flow-n-A-neg} for ${\sl cKP}_{R,M}$ hierarchy\foot{The 
flow $\d^{(-n)}_{E_k}$ for $k=1$ is excluded since
$\sum_{k=1}^{M+R} \d^{(-n)}_{E_k} = \d^{(-n)}_{\cA = \one}$ which vanishes
identially as explained in the previous section.} :
\be
\d^{(-n)}_{E_k} \equiv \pa/\pa \st{k}{t}\!\!{}_{-n}  \quad ,\quad
\cM^{(-n)}_{E_k} = \sum_{s=1}^n \P^{(-n-1+s)}_k D^{-1} \Psi^{(-s)}_k
\quad ,\;\; k=2,\ldots, M+R \; ,\;\; n=1,2,\ldots
\lab{ghost-k-neg}
\ee
which also span an infinite-dimensional Abelian algebra of flows commuting
with the isospectral flows. Then, following the steps of our construction in
ref.\ct{multi-comp-KP} we arrive at $(M+R)$-component constrained KP
hierarchy given in terms of $(M+R)(M+R-1)+1$ tau-functions
$\lcurl {\wti \t}_{ab} \rcurl$ obeying the corresponding $(M+R)\times(M+R)$
matrix Hirota bilinear identities (cf. \rf{HBI}). The latter tau-functions
are expressed in terms of the original single tau-function $\t$ and the
``negative'' flow generating (adjoint) eigenfunctions \rf{EF-sys-inverse}
in the original ordinary ${\sl cKP}_{R,M}$ hierarchy as follows:
\be
{\wti \t}_{11}={\wti \t}_{aa} = \t \quad ,\quad
{\wti \t}_{1a} = \t\, L_M ({\bar \vp}_{a}) \quad ,\quad 
{\wti \t}_{a1} = - \t\,{\bar \psi}_{a} 
\nonu
\ee
\be
{\wti \t}_{ab} = \vareps_{ab} \t\,\pai\( L_M ({\bar \vp}_{b}) {\bar \psi}_{a}\)
\quad ,\quad  a \neq b \;\; ,\;\; a,b = 2, \ldots M+R
\lab{tau-M+R-comp}
\ee

Let us recall that multi-component (matrix) KP hierarchies \rf{HBI}
contain various physically interesting nonlinear systems such as
Davey-Stewartson and $N$-wave systems, which now can be written entirely in
terms of objects belonging to ordinary one-component (constrained) KP hierarchy. 
For instance, the $N$-wave resonant system ($N=M(M+1)/2$) is given by:
\be
\pa_c f_{ij} = f_{ik} f_{kj} \quad ,\quad i\neq j\neq k \quad ,\quad
i,j,k=1,\ldots, M+1
\lab{N-wave-def}
\ee
\be
\pa_k \equiv \pa/\pa \st{k}{t}\!\!{}_1 \quad ,\quad
f_{1i} \equiv \P^{(1)}_{i-1} \quad ,\quad f_{i1} \equiv - \Psi^{(1)}_{i-1}
\nonu
\ee
\be
f_{ij} \equiv \vareps_{ij} \pai\(\P^{(1)}_{j-1}\Psi^{(1)}_{i-1}\)
\quad ,\quad  i \neq j \;\; ,\;\; i,j = 2, \ldots M+1
\lab{N-wave-id}
\ee

As a further example, we will demonstrate that the well-known
Davey-Stewartson system \ct{DS} arises as particular subset of symmetry flow
equations obeyed by any pair of adjoint eigenfunctions $\bigl(\P_i,\Psi_i\bigr)$
($i$=fixed) or $\bigl(L_M ({\bar \vp}_a), {\bar \psi}_a\bigr)$ ($a$=fixed).
In fact, for $\bigl(\P_i,\Psi_i\bigr)$ ($i$=fixed) this has already been
done in last ref.\ct{multi-comp-KP}. Here for simplicity we take
${\sl cKP}_{1,M}$ hierarchy (Eq.\rf{Lax-R-M} with $R=1$; the general case is
straightforward generalization of the formulas below) and consider a pair
of ``negative'' symmetry flow generating (adjoint) eigenfunctions, {\sl e.g.},
$\bigl(\phi \equiv L_M ({\bar \vp}_a), \psi \equiv {\bar \psi}_a\bigr)$
($a$=fixed), which obeys
the following subset of flow equations -- w.r.t. $\pa/\pa t_2$,
$\bpa \equiv \pa/\pa \st{a}{t}\!\!{}_{-1}$ and
$\pa/\pa \bt_2 \equiv \pa/\pa \st{a}{t}\!\!{}_{-2}$ 
(cf. Eqs.\rf{flow-n-A-neg-EF-inverse}) :
\be
\partder{}{t_2} \phi = \(\pa^2 + 2\sum_{i=1}^M \P_i \Psi_i\) \phi \quad ,\quad
\partder{}{t_2} \psi = 
- \(\pa^2 + 2\sum_{i=1}^M \P_i \Psi_i\) \psi 
\lab{EF-eqs-2-inverse}
\ee
\be
\bpa \phi = \cM^{(-1)}_1 (\phi) - \cL^{-1}(\phi)  \quad ,\quad
\bpa \psi = - \(\cM^{(-1)}_1\)^\ast (\psi) 
+ \(\cL^{-1}\)^\ast (\psi)
\lab{ghost-1-minus}
\ee
\be
\pa/\pa \bt_2 \phi = \cM^{(-2)}_1 (\phi) - \cL^{-2}(\phi)  \quad ,\quad
\pa/\pa \bt_2 \psi = - \(\cM^{(-2)}_1\)^\ast (\psi) 
+ \(\cL^{-2}\)^\ast (\psi)
\lab{ghost-2-minus}
\ee
where:
\be
\cM^{(-1)}_1 \equiv \phi D^{-1} \psi  \quad ,\quad
\cM^{(-2)}_1 \equiv \cL^{-1}(\phi) D^{-1} \psi +
\phi D^{-1} \(\cL^{-1}\)^\ast (\psi)
\lab{M-neg-1-2}
\ee
Using \rf{ghost-1-minus} we can rewrite Eqs.\rf{ghost-2-minus} as purely
differential equation w.r.t. $\bpa$ :
\be
\partder{}{\bt_2} \phi = 
\llb - \bpa^2 + 2\bpa\(\pai\(\phi \psi\)\)\rrb \phi
\quad ,\quad
\partder{}{\bt_2} \psi = 
\llb \bpa^2 - 2\bpa\(\pai\(\phi \psi\)\)\rrb \psi
\lab{ghost-2-minus-diff}
\ee
Now, introducing new time variable $T=t_2 - \bt_2$ and the short-hand notation
$Q \equiv \sum_{i=1}^M \P_i \Psi_i - 2(\phi \psi) -
2 \bpa \bigl(\pai (\phi \psi)\bigr)$, and subtracting
Eqs.\rf{ghost-2-minus-diff} from Eqs.\rf{EF-eqs-2-inverse}, we arrive at the
following system of $(2+1)$-dimensional nonlinear evolution equations:
\be
\partder{}{T} \phi = 
\Bigl\lb \h (\pa^2 + \bpa^2) + Q + 2 \phi \psi \Bigr\rb \phi
\lab{DS-dyn-1}
\ee
\be
\partder{}{T} \psi = 
- \Bigl\lb \h (\pa^2 + \bpa^2) + Q + 2 \phi \psi \Bigr\rb \psi
\lab{DS-dyn-2}
\ee
\be
\pa\bpa Q + (\pa +\bpa)^2 \(\phi \psi\) = 0
\lab{DS-nondyn}
\ee
which is precisely the standard Davey-Stewartson system \ct{DS} for the
``negative'' (adjoint) eigenfunction pair
$\bigl(\phi \equiv L_M ({\bar \vp}_a),\, \psi \equiv {\bar \psi}_a\bigr)$
($a$=fixed).

The construction in this Section allows us to employ the well-known \DB
techniques from ordinary one-component (scalar) KP hierarchies (full or
constrained) in order to obtain new soliton-like solutions of
multi-component (matrix) KP hierarchies (see last ref.\ct{multi-comp-KP}).
\mskp
{\bf 6. The Full Virasoro Algebra of Additional Symmetries}
\sskp
In refs.\ct{noak-addsym} we have constructed an essential modification to
the original Orlov-Schulman additional Virasoro symmetry flows
\ct{Orlov-Schulman} needed in the case of ${\sf cKP}_{R,M}$ reduced KP
models \rf{Lax-R-M} for $n \geq 0$, {\sl i.e.}, for the Borel subalgebra
(henceforth $\cL \equiv \cL_{R,M}$) :
\be
\d^V_n \cL = \Sbr{- \( M\cL^n\)_{-} + \cX_n}{\cL}
\lab{addsym-Vir}
\ee
or, equivalently:
\be
\d^V_n \cL = \Sbr{\( M\cL^n\)_{+} + \cX_n}{\cL} + \cL^n
\lab{addsym-Vir-1}
\ee
where $\d^V_n \simeq - L_{n-1}$ (in terms of standard Virasoro notations).
Here:
\be
\Sbr{\cL}{M}=\one  \quad ,\quad
M = \sum_{k\geq 1} k t_k \cL^{k-1} +
\sum_{j\geq 1} \( -j p_j(-[\pa]) \ln\t\) \cL^{-j-1}
\lab{M-def}
\ee
\be
\cX_n \equiv \sum_{i=1}^M \sum_{j=0}^{n-2} \bigl( j -\h (n-2) \bigr)
\cL^{n-2-j}(\P_i) D^{-1} \(\cL^j\)^\ast (\Psi_i)
\lab{X-n-def}
\ee
The presence of the additional terms $\cX_n$ in \rf{addsym-Vir}
is very crucial to ensure that the flows $\d^V_n$ preserve the constrained
form of the pertinent pseudo-differential Lax operator \rf{Lax-R-M}.
The ordinary Orlov-Schulman flows 
$\d_n^{OS} \cL = \Sbr{-\( M\cL^n\)_{-}}{\cL}$ do not define symmetries for 
constrained ${\sf cKP}_{R,M}$ hierarchies.

The action of $\d^V_n$-flows on the pertinent (adjoint) eigenfunctions reads
accordingly (for $n\geq 0$) :
\be
\d^V_n \P_i = \llb \( M\cL^n\)_{+} + \cX_n \rrb (\P_i) + 
{n\o 2} \cL^{n-1}(\P_i)
\lab{Vir-flow-EF-i}
\ee
\be 
\d^V_n \Psi_i = - \llb \( M\cL^n\)^\ast_{+} + \cX^\ast_n \rrb (\Psi_i) + 
{n\o 2} \(\cL^{n-1}\)^\ast (\Psi_i)
\lab{Vir-flow-adj-EF-i}
\ee
Similarly, for the (adjoint) eigenfunctions entering the inverse Lax powers
we find from ~$\d^V_n \cL^{-1} = \Sbr{- \( M\cL^n\)_{-} + \cX_n}{\cL^{-1}} $ 
and Eqs.\rf{zero-eqs} (for $n\geq 0$) :
\be
\d^V_n L_M ({\bar \vp}_a) = \llb \( M\cL^n\)_{+} + \cX_n \rrb 
\bigl( L_M ({\bar \vp}_a)\bigr)  \quad ,\quad
\d^V_n {\bar \psi}_a = 
- \llb \( M\cL^n\)^\ast_{+} + \cX^\ast_n \rrb ({\bar \psi}_a)
\lab{Vir-flow-EF-a}
\ee

Here we want to extend the above construction to cover the case of the full 
Virasoro algebra of additional symmetries. For the negative flows we must
therefore find the appropriate additional terms $\cX_{(-n)}$ :
\be
\d^V_{-n} \cL = \Sbr{- \( M\cL^{-n}\)_{-} + \cX_{(-n)}}{\cL}
\lab{addsym-Vir-minus}
\ee
or, equivalently:
\be
\d^V_{-n} \cL = \Sbr{\( M\cL^{-n}\)_{+} + \cX_{(-n)}}{\cL} + \cL^{-n}
\lab{addsym-Vir-minus-1}
\ee
so that the consistency condition \rf{consist-cond} is satisfied.
Using again the pseudo-differential operator identities 
\rf{pseudo-diff-id} and taking into account the relevant formulas
for negative Lax powers \rf{L-minus-N} we obtain the following explicit 
expressions for $\cX_{(-n)}$ :
\be
\cX_{(-n)} = \sum_{a=1}^{M+R} \sum_{j=0}^n \bigl({n\o 2}-j\bigr) 
\cL^{-(n-j)} \bigl( L_M ({\bar \vp}_a)\bigr) D^{-1} 
\(\cL^{-j}\)^\ast ({\bar \psi}_a) 
\lab{X-n-minus-def}
\ee
The consistency of the negative flow definitions \rf{addsym-Vir-minus-1} with 
$\cX_{(-n)}$ as in Eq.\rf{X-n-minus-def} crucially depends on the relations 
\rf{zero-eqs}.

The flows $\d^V_{-n}$ act on the constituent (adjoint) eigenfunctions of $\cL$
as:
\be
\d^V_{-n} \P_i = \llb\( M\cL^{-n}\)_{+} + \cX_{(-n)}\rrb (\P_i)  \quad ,\quad
\d^V_{-n} \Psi_i = - 
\llb\( M\cL^{-n}\)^\ast_{+} + \cX_{(-n)}^\ast\rrb (\Psi_i)
\lab{addsym-Vir-minus-EF}
\ee
and similarly on the (adjoint) eigenfunctions $L_M ({\bar{\vp}}_a)$, 
${\bar {\psi}}_a$ entering the inverse powers of $\cL$ :
\be
\d^V_{-n} \bigl( L_M ({\bar{\vp}}_a)\bigr) = 
\llb\( M\cL^{-n}\)_{+} + \cX_{(-n)}\rrb \bigl( L_M ({\bar{\vp}}_a)\bigr)
- \bigl({n\o 2}+1\bigr) \cL^{-(n+1)} \bigl( L_M ({\bar{\vp}}_a)\bigr)
\lab{addsym-Vir-minus-EF-inverse}
\ee
\be
\d^V_{-n} {\bar {\psi}}_a  = - 
\llb\( M\cL^{-n}\)^\ast_{+} + \cX_{(-n)}^\ast\rrb ({\bar {\psi}}_a)
- \bigl({n\o 2}+1\bigr) \(\cL^{-(n+1)}\)^\ast ({\bar {\psi}}_a)
\lab{addsym-Vir-minus-adj-EF-inverse}
\ee

Let us now consider the commutator of the Virasoro flows
$\d^V_{n} \simeq - L_{n-1}$ and $\d^V_{m} \simeq - L_{m-1}$ acting on $\cL$
(cf. Eq.\rf{commut-flows}) where $(n,m)$ are arbitrary non-negative or 
negative indices:
\be
\Sbr{\d^V_{n}}{\d^V_{m}} = 
\d^V_{n} \Bigl( -\( M\cL^{m}\)_{-} + \cX_{m}\Bigr) -
\d^V_{m} \Bigl( -\( M\cL^{n}\)_{-} + \cX_{n}\Bigr)
- \Sbr{ -\( M\cL^{n}\)_{-} + \cX_{n}}{-\( M\cL^{m}\)_{-} + \cX_{m}}
\lab{commut-n-m}
\ee
Using the identity:
\br
\d^V_n \( M\cL^m\)_{-} - \d^V_m \( M\cL^n\)_{-} =
-(n-m) \( M\cL^{n+m-1}\)_{-} \nonu \\
-\Sbr{\( M\cL^n\)_{-}}{\( M\cL^m\)_{-}} + {\Sbr{\cX_n}{M\cL^m}}_{-} -
{\Sbr{\cX_m}{M\cL^n}}_{-}
\lab{rel-n-m}
\er
the r.h.s. of Eq.\rf{commut-n-m} can be rewritten in the form:
\be
(n-m) \( M\cL^{n+m-1}\)_{-} + \d^V_n \cX_m - {\Sbr{\( M\cL^n\)_{+}}{\cX_m}}_{-}
- \d^V_m \cX_n + {\Sbr{\( M\cL^m\)_{+}}{\cX_n}}_{-} - \Sbr{\cX_n}{\cX_m}
\lab{commut-n-m-rhs}
\ee
Now, employing the pseudo-differential identities \rf{pseudo-diff-id} it
is easy to show, taking into account \rf{Vir-flow-EF-i}--\rf{Vir-flow-EF-a}
and \rf{addsym-Vir-minus-EF}--\rf{addsym-Vir-minus-adj-EF-inverse},
that the sum of all terms in \rf{commut-n-m-rhs} involving $\cX_{n,m}$ 
yield:
\be
\d^V_n \cX_m - {\Sbr{\( M\cL^n\)_{+}}{\cX_m}}_{-}
- \d^V_m \cX_n + {\Sbr{\( M\cL^m\)_{+}}{\cX_n}}_{-} - \Sbr{\cX_n}{\cX_m}
= -(n-m) \cX_{n+m-1}
\lab{X-terms}
\ee
Thus, we have verified the closure of the full Virasoro algebra of additional
symmetries without central extension:
\be
\Sbr{\d^V_{n}}{\d^V_{m}} = -(n-m) \d^V_{n+m-1}
\lab{full-Vir-alg}
\ee

\mskp
{\underline{\bf Outlook}}
In a subsequent paper we will generalize the present
construction of additional symmetries to the case of supersymmetric
integrable hierarchies. We will continue the derivation and study of
properties of new soliton-like solutions of multi-component KP hierarchies
obtained via standard \DB methods for ordinary one-component KP models which
has been already initiated in the last ref.\ct{multi-comp-KP}. 
In a forthcoming more detailed paper we will systematically study the
construction of additional (loop-algebra and Virasoro) symmetries within 
a generalized Drinfeld-Sokolov formalism both in ordinary and
supersymmetric integrable systems of KP type. 
Also we will relate the algebraic dressing method to Sato
pseudo-differential operator approach.
\sskp
{\underline{\bf Acknowledgements}}
H.A. is partially supported by NSF ({\sl PHY-9820663}) and 
J.F.G. is partially supported by CNPq and Fapesp (Brazil).
Three of us (H.A., E.N. and S.P.) gratefully acknowledge support from NSF
grant {\sl INT-9724747}. E.N. and S.P. are partially supported by Bulgarian
NSF grant {\sl F-904/99}.

\end{document}